\documentclass[10pt, conference]{IEEEtran}
\IEEEoverridecommandlockouts
\newif\ifARXIV
\ARXIVtrue

\ifARXIV
\else
\fi

\usepackage{cite}
\usepackage{amsmath,amssymb,amsfonts}
\usepackage{algorithmic}
\usepackage{graphicx}
\usepackage{textcomp}

\usepackage{multirow}

\usepackage[hyphens]{url}
\ifARXIV
\usepackage{hyperref}
\fi

\usepackage{listings}

\usepackage{caption}
\begin{document}

\title{General Purpose Atomic Crosschain Transactions
% \thanks{Both authors are full-time employees of ConsenSys Software}
}

\author{
    \IEEEauthorblockN{
    	Peter Robinson\IEEEauthorrefmark{1}\IEEEauthorrefmark{2},
    	Raghavendra Ramesh\IEEEauthorrefmark{1} 
         } 
    \IEEEauthorblockA{\IEEEauthorrefmark{1}ConsenSys Software R\&D}
    \IEEEauthorblockA{\IEEEauthorrefmark{2}School of Information Technology and Electrical Engineering, University of Queensland, Australia}
    \IEEEauthorblockA{
    	peter.robinson@consensys.net, 
    	raghavendra.ramesh@consensys.net
	}
}

% if ARXIV ********************************************************************
\ifARXIV
\maketitle

\else
\IEEEoverridecommandlockouts
\IEEEpubid{\makebox[\columnwidth]{978-1-6654-3924-4/21/\$31.00~\copyright2021 IEEE \hfill} \hspace{\columnsep}\makebox[\columnwidth]{ }}

\maketitle

\IEEEpubidadjcol

% fi ARXIV ********************************************************************
\fi

% Add page numbers
%\thispagestyle{plain}
%\pagestyle{plain}

% Include:
% Background: A simple opening sentence or two placing the work in context. 
% Aims: One or two sentences giving the purpose of the work. 
% Method(s): One or two sentences explaining what was done. 
% Results: One or two sentences indicating the main findings. 
% Conclusions: One sentence giving the most important consequence of the work.
\begin{abstract}
The General Purpose Atomic Crosschain Transaction protocol allows composable programming across multiple Ethereum blockchains. It allows for inter-contract and inter-blockchain function calls that are both synchronous and atomic: if one part fails, the whole call execution tree of function calls is rolled back. The protocol operates on existing Ethereum blockchains without modification. It works for both public permissioned and consortium blockchains.
\ifARXIV
 Additionally, the protocol is expected to work across heterogeneous blockchains other than Ethereum.
\fi
This paper describes the protocol, analyses it in terms of Gas usage and \textit{Finalised Block Periods} for three scenarios: reading a value from one blockchain to another, writing a value from one blockchain to another, and a trade finance system involving five contracts on five blockchains with a complex call execution tree, and provides an initial security analysis that shows that the protocol has \textit{Safety} and \textit{Liveness} properties. 
\end{abstract}


%%%%%%%%%%%%%%%%%%%%%%%%%%%%%%%%%%%%%%%%%%%%%%%%%%%%%%%
%%%%%%%%%%%%%%%%%%%%%%%%%%%%%%%%%%%%%%%%%%%%%%%%%%%%%%%
\section{Introduction}
\label{sec:introduction}
The General Purpose Atomic Crosschain Transaction (GPACT) protocol is a blockchain technology that allows function calls across blockchains that either updates state on all blockchains or discards state updates on all blockchains. 
% if ARXIV ********************************************************************
\ifARXIV
The function calls can return values across blockchains. 
% fi ARXIV ********************************************************************
\fi
The protocol enables applications to access information and utilise functionality that resides on one blockchain from other blockchains. Previous atomic crosschain protocols~\cite{hashtimelock, lightning2016, interledger, dogethereum} only offer atomic asset swaps. We use the term \textit{General Purpose} to mean that this protocol is not tied to a specific application. Almost any application built on the Ethereum Virtual Machine can be created as a crosschain application using this protocol.

Supply chain logistics and finance are key use-cases of blockchain technologies. Trade Finance is the combination of these two systems in which financiers operate on one blockchain and supply chain participants operate on another blockchain. The GPACT protocol has been described with this use-case in mind. The protocol is however generally applicable to any use-case that involves function calls across blockchains.

\begin{figure}[b]
  \includegraphics[width=\linewidth]{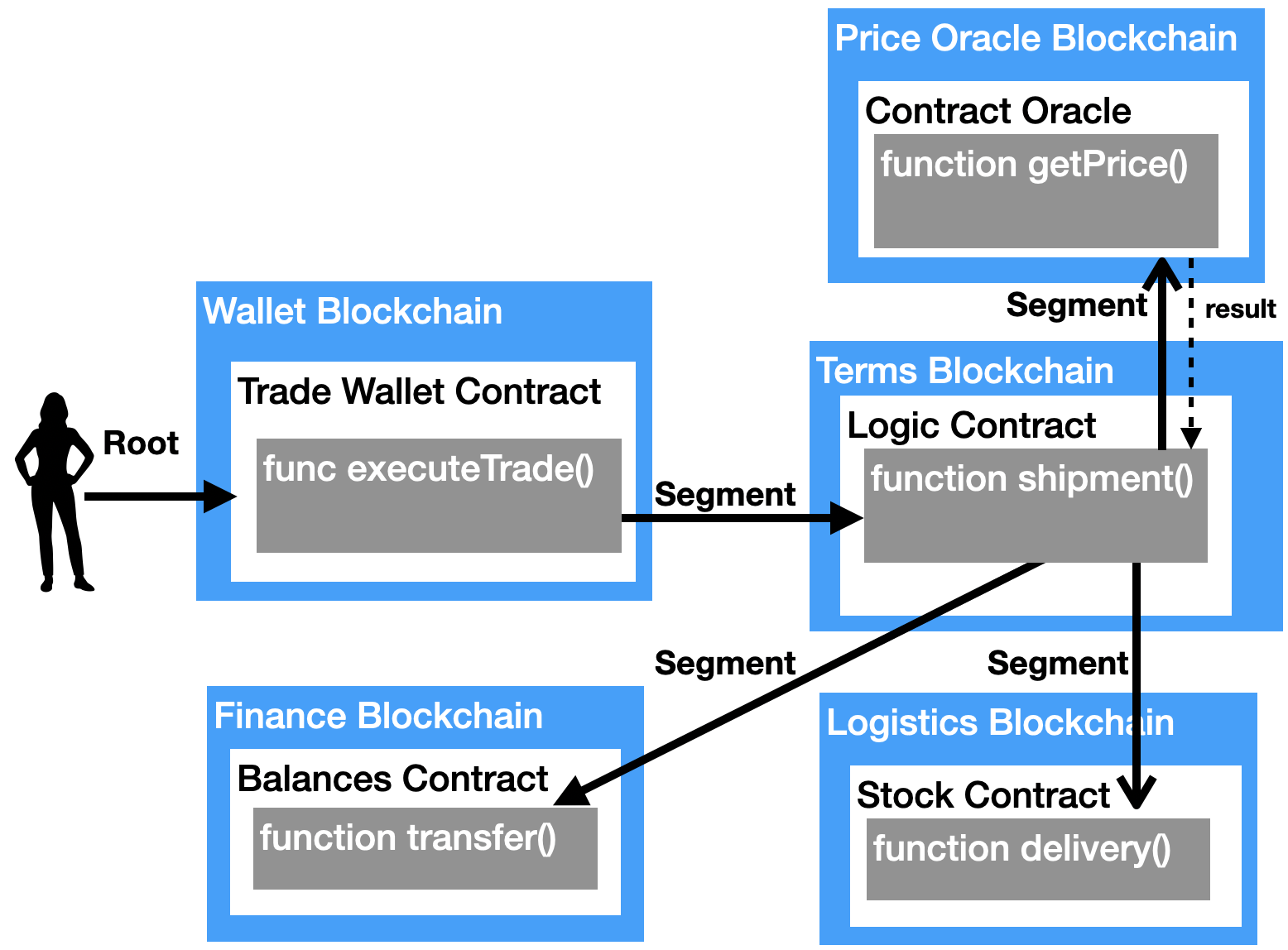}
  \caption{Trade Finance Using Atomic Crosschain Transactions}
  \label{fig:usage}
\end{figure}

Figure~\ref{fig:usage} shows a logical representation of a crosschain call execution tree for the Trade Finance use-case. A trade finance application creates a crosschain function call that goes across five contracts on five blockchains to execute a trade for a shipment of goods. The Root Transaction executes the entry point function, the \texttt{executeTrade} function in the \texttt{Trade Wallet} contract on the \texttt{Wallet} blockchain. The \texttt{Trade Wallet} contract could be a multi-signatory wallet that parties to a shipment have to submit a transaction to, indicating that they agree a shipment for a certain quantity of goods has been made and should be paid for. The \texttt{executeTrade} function calls the \texttt{shipment} function in the \texttt{Logic} contract on the \texttt{Terms} blockchain to determine the price that should be paid and to affect the transfer of stock and payment. The \texttt{shipment} function calls the \texttt{getPrice} function on the \texttt{Oracle} contract on the \texttt{Price Oracle} blockchain to determine the price that should be paid for the goods, then calls the \texttt{transfer} function on the \texttt{Balances} contract on the \texttt{Finance} blockchain to affect the payment, and finally calls the \texttt{delivery} function on the \texttt{Stock} contract on the \texttt{Logistics} blockchain to register the changed ownership of the goods.

It could be argued that some of the contracts could exist on the one blockchain, thus reducing the need for crosschain transactions. However, the \texttt{Finance} blockchain and the \texttt{Logistics} blockchain in particular could be consortium blockchains involving different participants. The \texttt{Price Oracle} blockchain could be operated by a consortium that charged for access to the information. Government regulators could require the logic on the \texttt{Terms} blockchain visible to them, but the participants in the trade wallet on the \texttt{Wallet} blockchain may wish to remain anonymous. A crosschain transaction capability is needed to meet these requirements.

The contributions of this paper are a description and analysis of a protocol that provides atomic function calls across blockchains. 
Source code and documentation describing how to reproduce the results are available on github.com~\cite{gpact_github}. 

This paper is organised as follows: the \textit{Related Works} section describes four crosschain consensus methodologies for communicating Ethereum Events across blockchains, how these techniques can be used by this protocol and how they relate to the GPACT protocol. Next the \textit{Protocol Description} section explains the protocol. The \textit{Security Analysis} section analyses the Safety and Liveness of the protocol. The \textit{Evaluation and Discussion} section presents and analyses the gas usage and latency of the protocol. 
% if ARXIV ********************************************************************
\ifARXIV
The \textit{Crosschain for Heterogeneous Blockchains} section explains how the protocol can be extended to allow crosschain function calls that include non-Ethereum blockchains.
% fi ARXIV ********************************************************************
\fi
The limitations of this protocol are outlined in the \textit{Limitations} section.

%%%%%%%%%%%%%%%%%%%%%%%%%%%%%%%%%%%%%%%%%%%%%%%%%%%%%%%
%%%%%%%%%%%%%%%%%%%%%%%%%%%%%%%%%%%%%%%%%%%%%%%%%%%%%%%
\section{Related Works}
\label{ref:trust}
The protocol relies on transferring information between blockchains using Ethereum Events. Ethereum~\cite{wood2016a} is a blockchain platform that allows users to deploy and execute computer programs known as Smart Contracts. Transactions can be used to deploy contracts, execute function calls on the contracts, and to transfer Ether between accounts. Transactions in Ethereum can programmatically generate log events that are stored in transaction receipts. The transaction receipts for all transactions in a block are stored in a modified Merkle Patricia Tree. The root hash of this tree is stored in the Ethereum block header. The log event information includes: the address of the contract that emitted the event, an identifier known as a \textit{topic} that specifies the type of event that is emitted, and a data blob containing the encoded event parameters. This ability to programmatically produce events can be used to produce information on a source blockchain that can be consumed on a destination blockchain. This section describes four techniques for communicating events across blockchains. 

\subsection{Direct Signing}
The direct signing technique requires validators for each blockchain to be registered on all other blockchains. A threshold number of validators sign event information contained in transaction receipts. Contract code on remote blockchains trust the signed event information because a threshold number of signers registered in a contract on the remote blockchain have signed the event information. Cosmos Inter-Blockchain Communications (IBC)~\cite{cosmos2016} uses this technique with a central hub blockchain to provide crosschain communications and Nissl et al.~\cite{nissl2020crossblockchain} describe a crosschain framework for contract invocations across blockchains using this technique. 

\subsection{Block Header Transfer}
\label{ref:bhtransfer}
Block Header Transfer~\cite{robinson-consensus-crosschain} techniques rely on relayers reading block headers from a source blockchain and submitting the block headers to a destination blockchain. Contract code on destination blockchains trust event information because a threshold number of relayers registered in a contract on the destination blockchain have signed the block header that the transaction that generated the event was included in, plus the event is proven to have been produced by a transaction in the block using a Merkle Proof. This is the technique used by Clearmatics' Ion project~\cite{Clearmatics2018c}. 

An advantage of this scheme is that relayers only have to sign a single block header for all crosschain transactions in a block, rather than having to sign each crosschain transaction. The main disadvantage is that transactions are needed to submit the signed block header to the destination blockchain in addition to the transaction executing the application logic. This increases the latency of the protocol using the technique. 
% if ARXIV ********************************************************************
\ifARXIV
An additional issue is that the relayers need to have permission to submit transactions on the destination blockchain, which in permissioned blockchain usages may not be the case. 
% fi ARXIV ********************************************************************
\fi

\subsection{Block Header Transfer via a Coordinating Blockchain}
Early proposals for Ethereum 2 had multiple execution shards post root hashes called \textit{Crosslinks} to the Beacon Chain~\cite{ethereum2-beacon-chain}. The Crosslinks for all shards would be available to all shards in the next block. This mechanism of cross-linking would allow events from a source shard to be trusted on a destination shard by presenting a Merkle Proof to demonstrate that the event was generated by a transaction on the source shard, similarly to the block header transfer technique described above.

Hyperservice's Network Status Blockchain's~\cite{hyperservice} holding of crosschain state in a Merkle Tree is similar technique to Ethereum 2, posting signed block headers to a coordination blockchain and then publishing root information from the coordination blockchain to all blockchains. In this way, an event as a result of a transaction on any blockchain can be trusted on any other blockchain.

Ethereum 2 has been especially constructed to have crosslinks available for the next shard block. However, general purpose systems are unlikely to be able to achieve such low latency. This increased latency between when a transaction occurs on a source blockchain and when the event information from the transaction can be consumed on a destination blockchain is the key downsize of this approach.

% if ARXIV ********************************************************************
\ifARXIV
An issue with this type of construction is that the attack surface of the system expands to include all blockchains. If any of the blockchains become untrusted by a particular blockchain, there is no ability for that blockchain to selectively block the trusting of information from the untrusted source blockchain.
% fi ARXIV ********************************************************************
\fi

\subsection{Simple Payment Verification}
Simplified Payment Verification~\cite{nakamoto2008, btc-relay, robinson-consensus-crosschain} can be used to transfer block headers from a source blockchain such that they are trusted on a destination blockchain by virtue of the PoW difficulty of the source blockchain. The destination blockchain needs to be able to programmatically prove the PoW calculation, and hence is practically limited to source blockchains that have a simple PoW calculation and destination blockchains that have an expressive enough contract programming language to allow the calculation algorithm to be created. Additionally, the source blockchain's PoW difficulty should be high enough such that attackers can not mount a 51\% attack, thus forking the block headers being accepted by the destination blockchain.

\subsection{Summary of Event Transfer Techniques}
Four techniques have been described for transferring programmatically generated events from a source blockchain to a destination blockchain such that the events will be trusted. None of the techniques provide a mechanism for atomic updates due to function calls across blockchains. 

\subsection{Atomic Crosschain Techniques}
Multiple crosschain and cross-shard techniques have been proposed that require changes to blockchain client software~\cite{robinson2019b, chainspace, omniledger}, and are hence not  generally applicable as organisations are unwilling to make changes to existing blockchain protocols and software. Atomic swap protocols~\cite{hashtimelock, lightning2016, interledger, dogethereum} only offer atomic asset swaps. Cross-Hyperledger Fabric channel~\cite{cross-channel} protocols have been proposed, but don't address the distributed trust requirements of crosschain. Crosschain deals~\cite{crosschain-deals} that require specialised set-up for each transaction have been proposed. They don't allow for complex crosschain programs to be composed using standard programming techniques. 

The GPACT protocol is similar to the Tree Two Phase Commit~\cite{tree2pc} protocol, Start and Segment being analogous to 2PC Prepare phase and Root and Signalling being analogous to the 2PC Commit and Forget phase. The GPACT protocol provides blockchain specific mechanisms for how to commit to a call execution tree, how to execute parts of the call execution tree and how to do locking, that the Tree Two Phase Commit protocol does not describe.

The techniques described in this paper build on the earlier work on Ethereum 2 cross-shard function calls~\cite{robinson-cross-shard}, and an open access version of this paper~\cite{robinson-layer2-arxiv} written prior to implementation.

%%%%%%%%%%%%%%%%%%%%%%%%%%%%%%%%%%%%%%%%%%%%%%%%
\section{Protocol Description}
\label{sec:methodology}

\subsection{Overview}
Applications use the GPACT protocol when they wish to execute a function call who's call execution tree goes across functions in contracts on more than one blockchain. The first step is to determine the parameter values that need to be passed into entry point functions on each blockchain. Then segments of the call execution tree are executed, from the leaves of the call execution tree to the root. Contracts are locked if a call segment updates the state of a contract. These provisional updates are applied to all locked contracts if the overall crosschain transaction is successful, and discarded if not.

\subsection{Call Execution Tree Simulation}
The first step to execute the protocol is to determine the expected parameter values for entry point function calls for contracts on blockchains. The application needs to fetch state from the contracts and then execute a simulation of the contract code. For example, in Figure~\ref{fig:usage}, the value that is expected to be returned by the \texttt{getPrice} function needs to be fetched from the \texttt{Oracle} contract on the \texttt{Price Oracle} blockchain. The expected parameter values for each function call across the call execution tree can then be determined given the \texttt{buyer}, \texttt{seller}, and the \texttt{quantity} parameters to be passed into the \texttt{executeTrade} function in the \texttt{Trade Wallet} contract on the \texttt{Wallet} blockchain.

%From Raghavendra:
%To execute a crosschain contract function call transaction, we need to execute its individual transactions on respective blockchains. An individual transaction is again a contract function call, and to execute it we need to provide values to its parameters. The initiating or the originating transaction provides the values to the parameters of the top function, but not the parameters to the function calls. Since the originating node is operating on all blockchains, it has the source code of all the callee (recursively) contract functions and can simulate (locally execute without persisting the state) and determine the parameter values for all the nested callee functions. Hence we fix the originating node to fetch the states of contracts across all respective blockchains, lock the contracts across all blockchains, and simulate the crosschain function call fully, and determine the parameter values required for the individual sub transactions.

\subsection{Crosschain Control Contract}
The next steps of the protocol use Crosschain Control Contracts to manage parts of crosschain function calls. An instance of this contract is deployed to each blockchain that will participate in crosschain function calls. The address of the contract on each blockchain and a set of signers for each blockchain are registered with Registrar Contracts on all blockchains. This allows events emitted by this contract to be trusted as being from a Crosschain Control Contract. The following sections describe the main functions of the contract. 
\begin{figure*}
\begin{center}
  \includegraphics[width=0.9\linewidth]{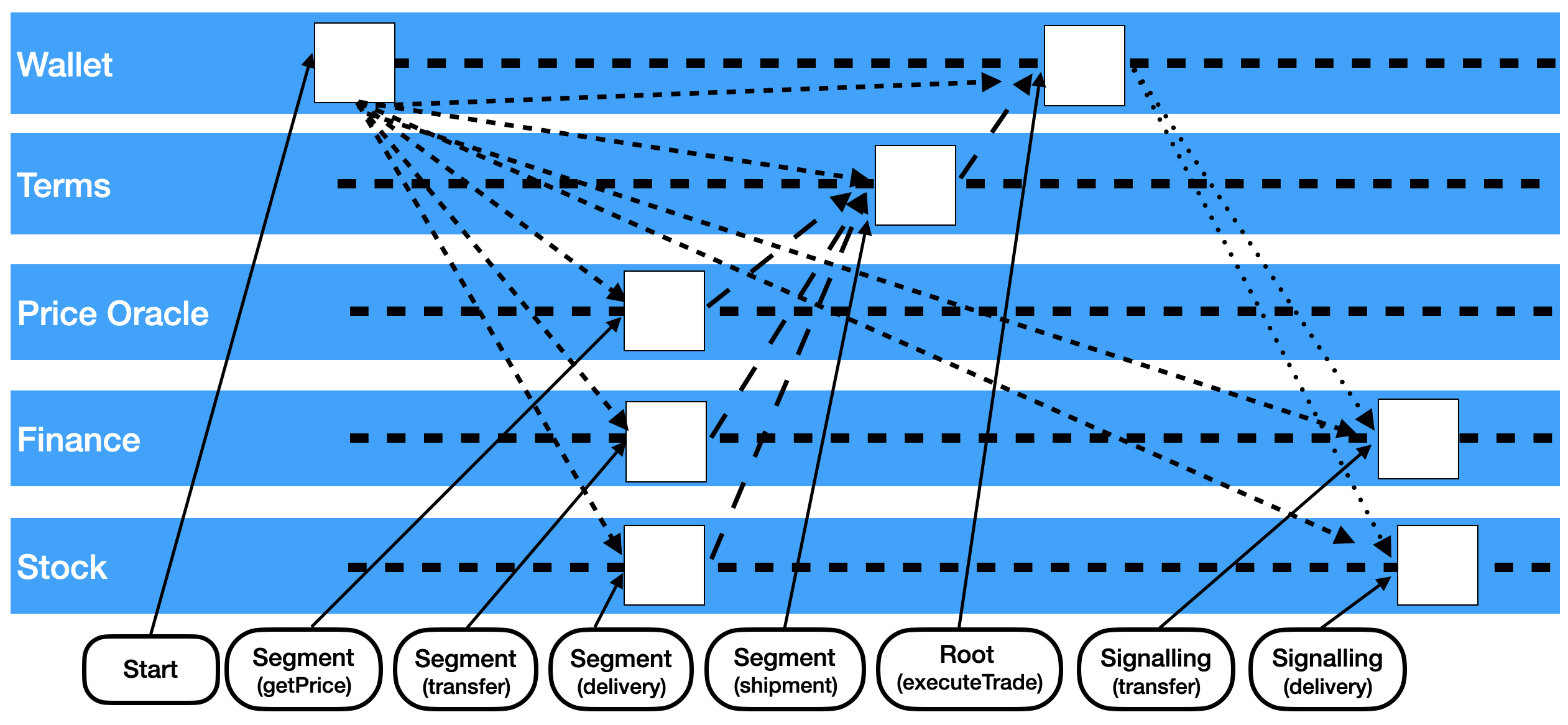}
  \caption{Example Sequence of Function Calls}
  \label{fig:sequence}
\end{center}
\end{figure*}

\subsubsection{Start}
The \textit{Start} function is called on the blockchain which contains the entry point function call for the call execution tree, the Root Blockchain. The function registers the account that will submit all of the transactions on all of the blockchains, the call execution tree of the crosschain function call including expected function parameter values, a time-out in terms of block time stamp of the Root Blockchain, and a crosschain transaction identifier. This information is emitted in an event known as the \textit{Start Event}. The transaction that submits the Start function call is known as the \textit{Start Transaction}.

\subsubsection{Segment}
The \textit{Segment} function is called to request a function on a contract be called as part of a crosschain function call. The signed Start Event and an indicator of where the function call lies in the call execution tree are submitted as parameters to prove that this function is part of the crosschain function call. Additionally, a set of signed \textit{Segment Events} containing function call return results are passed in to prove that subordinate function calls on other blockchains have been called and have returned certain results. 

When the application contract code executes, if a crosschain function call is encountered, the application code calls a function \textit{CrossCall} in the Crosschain Control Contract, passing in the actual parameters of the function call. The  CrossCall function compares the actual parameter values with the expected parameter values as specified in the Start Event. The function returns results that have been indicated should be returned in Segment Event for the function call. Section \ref{sec:locking} describes contract locking, which is used if state updates occur.

A Segment Event is emitted to publish the return result or error result of the function, and the list of locked contracts. The transactions that submit the Segment function calls are known as \textit{Segment Transactions} and the blockchains they execute on are termed \textit{Segment Blockchains}.

\subsubsection{Root}
The \textit{Root} function is called on the Root Blockchain to call the entry point function call for the call execution tree. The Root function has similar parameters to the Segment function, taking a signed Start Event and a set of signed Segment Events. Similar to the Segment function, expected and actual parameter checking is completed along with contract locking.

If a Root function completes successfully, any locked contracts on the Root Blockchain are unlocked and provisional state updates are committed. A \textit{Root Event} is emitted indicating that all provisional updates on blockchains should be committed. The Root function emits a Root Event indicating that all updates on all other blockchains should be discarded if any of the Segment functions returned error results or if an error occurs while executing the entry point function. 

If the time stamp of the most recent block on the Root Blockchain is after time-out, then any account can submit a transaction that calls the Root function to cancel the crosschain function call. In this situation a Root Event is emitted that indicates that all updates on all other blockchains should be discarded.

\subsubsection{Signalling}
The \textit{Signalling} function is called on blockchains that have updates that need to be committed or aborted. The signed Root Event and signed Segment Events for the blockchain are passed in as parameters. Calling the function requests that contracts on a blockchain locked due to a state update due to a Segment function call be unlocked and the state updates be either committed or aborted depending on the information in the Root Event. A Signalling Event is emitted to indicate that the contract has been unlocked.

\subsection{Call Execution Trees and Reverse Order Execution}
Figure~\ref{fig:usage} shows a logical representation of a crosschain call in which an application creates a crosschain function call that goes across five blockchains. The set of Crosschain Control Contract function calls that should be executed to affect this is shown in Figure~\ref{fig:sequence}.

Walking through the call sequence diagram in Figure~\ref{fig:sequence}:
\begin{enumerate}
\item A Start Transaction is submitted to the Wallet blockchain, causing the Start Event to be emitted. This transaction registers the call execution tree for the crosschain transaction. 
\item Segment Transactions are submitted to the Price Oracle, Finance, and Stock blockchains to execute the leaf parts of the call execution tree; the functions \texttt{getPrice}, \texttt{transfer} and \texttt{delivery} functions. Segment Events are emitted as a result of these transactions. Using one of the techniques described in Section~\ref{ref:trust}, the Start Event that is submitted as part of the Segment Transactions is trusted by the Crosschain Communications contracts on the Price Oracle, Finance, and Stock blockchains, allowing these contracts to know the call execution tree that has been committed to.  
\item A Segment Transaction is submitted to the Terms blockchain. This executes the call next up the call execution tree from the leaves, the \texttt{shipment} function. A Segment Event is emitted as a result of this transaction. Using one of the techniques described in Section~\ref{ref:trust}, the Start Event and Segment Events that are submitted as part of the Segment Transaction are trusted by the Crosschain Communications contracts on the Terms blockchain. In particular, the result of the \texttt{getPrice} function contained in the Segment Event from the Price Oracle blockchain can be trusted and used when the \texttt{shipment} function calls the \texttt{getPrice} function. A Segment Event is emitted. 
\item The Root Transaction is submitted to the Wallet blockchain. This executes the root part of the call execution tree and emits a Root Event. Using one of the techniques described in Section~\ref{ref:trust}, the Start Event, and the Segment Event from the Terms blockchain that are submitted as part of the Root Transaction are trusted by the Crosschain Communications contracts on the Wallet blockchain. This allows the Crosschain Communications contract to know that the transaction has not timed out and the function call to be executed as part of the entry point function call (from the Start Event), and that the call to the \texttt{shipment} and the rest of the call execution tree has executed without failure. 
\item A Signalling Transaction is submitted to the Finance and the Logistics blockchain to commit the provisional state updates. Signalling transactions are not required on the Price Oracle or the Terms blockchains as neither of these blockchains update state as part of the crosschain function call.
\end{enumerate}

% if ARXIV ********************************************************************
\ifARXIV
It should be noted that Figure~\ref{fig:sequence} makes some assumptions. Note that these are not limitations of the protocol. They are merely limitations of the diagrammatic representation of an example usage.
\begin{itemize}
\item The diagram only shows blocks being produced when transactions are present. Many consensus algorithms produce blocks irrespective of whether there are transactions present.
\item The diagram shows block production times synchronised across blockchains. This is typically not the case.
\item It has been assumed an instant finality consensus algorithm such as IBFT2~\cite{ibft2} has been used on all blockchains. Signed events or signed block headers can only be transferred once a block is deemed final. 
\item A parallel execution engine is assumed, where all Segments Transactions at the same level of call execution tree can be executed simultaneously.
\end{itemize}

\fi
% fi ARXIV ********************************************************************

\subsection{Contract Call Flow}
\begin{figure}
  \includegraphics[width=\linewidth]{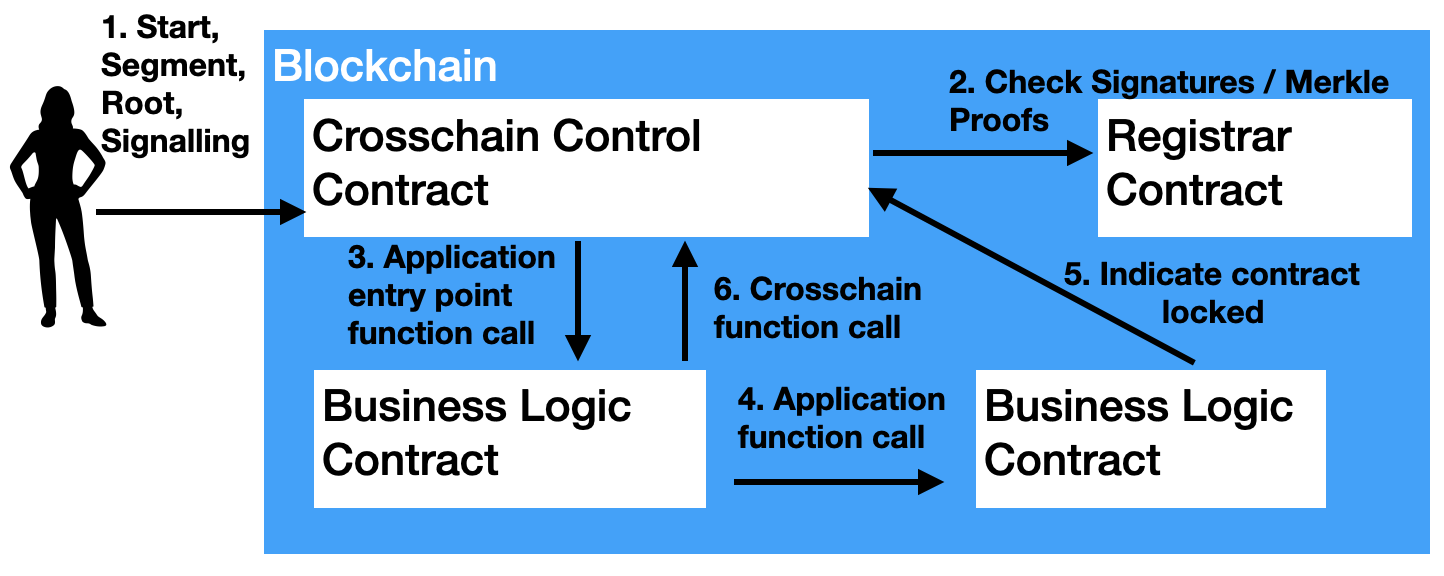}
  \caption{Example Contract Call Flow}
  \label{fig:calls}
\end{figure}

Figure~\ref{fig:calls} shows the call flow between contracts to facilitate the protocol. Walking through the steps in the diagram:
\begin{enumerate}
\item The application submits Start, Segment, Root, and Signalling function calls to the Crosschain Control contract. 
\item The Segment, Root, and Signalling function parameters include signatures or Merkle Proofs to prove events from other blockchains are valid. The Registrar contract is used to check the validity of this event information. 
\item The Segment and Root function calls execute functions on business logic contracts. 
\item These business logic contracts may in turn call other business logic contracts. 
\item So Segment Events can emit the list of locked contracts, Business Logic contracts need to tell the Crosschain Control contract when items in a contract are locked. 
\item Business logic contracts may include crosschain calls. To check that there is linkage between Root and Segment functions, business logic contracts 
pass the blockchain id, contract address and parameters of the crosschain call to the Crosschain Control contract. It checks that the actual values match the expected values from the Start Event.
\end{enumerate}

\subsection{Locking}
\label{sec:locking}
Business logic contracts hold the parts of their data that can be updated in a crosschain function call in lockable storage. State updated due to a crosschain call is held as provisional state updates. Provisional updates are committed or discarded based on Signalling Transactions. The algorithms for processing read requests is shown in Listing~\ref{listing:read}, write requests is shown in Listing~\ref{listing:write}, and signalling requests is shown in Listing~\ref{listing:signal}.

\begin{lstlisting}[
%  frame=single,
  basicstyle=\footnotesize\ttfamily,
  numbers=left,
stepnumber=1, 
  firstnumber=1,
  numberfirstline=true,
  numbersep=5pt,    
  xleftmargin=0.5cm,
  morekeywords={msg},
  label=listing:read,
  caption=Locking: Read Request Processing
]
If locked throw an error
Read from normal storage
\end{lstlisting}

\begin{lstlisting}[
%  frame=single,
  basicstyle=\footnotesize\ttfamily,
  numbers=left,
stepnumber=1, 
  firstnumber=1,
  numberfirstline=true,
  numbersep=5pt,    
  xleftmargin=0.5cm,
  morekeywords={msg},
  label=listing:write,
  caption=Locking: Write Request Processing
]
If locked throw an error
Check Crosschain Control Contract:
 is there an active crosschain call?
If not (normal single blockchain call)
   Write to normal storage
Else (this is a crosschain call)
  Lock the contract.
  Indicate in the Crosschain Control 
    Contract that this call is locking 
    items in the contract
  Write to provisional storage
\end{lstlisting}

\begin{lstlisting}[
%  frame=single,
  basicstyle=\footnotesize\ttfamily,
  numbers=left,
stepnumber=1, 
  firstnumber=1,
  numberfirstline=true,
  numbersep=5pt,    
  xleftmargin=0.5cm,
  morekeywords={msg},
  label=listing:signal,
  caption=Locking: Signalling Request Processing
]
For all items in the contract locked by the 
crosschain transaction identified by (root 
blockchain and crosschain transaction):
  If committing updates
    Apply updates from provisional storage
  Else (discarding updates}
    Delete provisional storage
  Unlock item
\end{lstlisting}

\subsection{Parallel and Serial Execution Engines}
The transaction sequence shown in Figure~\ref{fig:sequence} assumes a Parallel Execution Engine, that executes all Segment Transactions at a certain level of the call execution tree at the same time. This execution engine assumes that there is no part of the call execution tree that reads from a contract that is then later written to by another part of the call execution tree. Such parallel execution would risk failure as the Segment Transaction updating and locking the contract may have executed prior to the Segment Transaction to read from the contract. In this case, a Serial Execution Engine that executes the Segments in call sequence order should be used. A Parallel Execution Engine results in lower latency for a crosschain transaction than a Serial Execution Engine as Segment Transactions can be executed in parallel. Signalling transactions are executed in parallel for both execution engine types as there are no ordering restrictions.

\ifARXIV
\subsection{CallExecution Tree Limitation}
Entities can only submit crosschain transactions across blockchains that they have access to. For example, an enterprise could only execute the call execution tree shown in Figure~\ref{fig:usage} if they could submit transactions to each of the blockchains that the call execution tree is across. 
\fi

%%%%%%%%%%%%%%%%%%%%%%%%%%%%%%%%%%%%%%%%%%%%%%%%%%%%%%
\section{Security Analysis}
The property of \textit{atomicity} in this protocol can be split into the following safety and liveness properties. 

It has been assumed that:
\begin{description}
\item[A1] The event signing or the block header transfer mechanism is trusted.
\item[A2] Root transaction will eventually (after a finite number of steps) be submitted either by the application or by others.
\item[A3] Signalling transactions will eventually (after a finite number of steps) be submitted either by the application or by others.
\item[A4] The number of Byzantine nodes on each blockchain is less than the limit imposed by its underlying consensus protocol.
\end{description}

\subsection{Safety}
\textbf{Claim}: Suppose that the protocol has finished executing a crosschain transaction \texttt{c}. Then:
\begin{itemize}
\item If \texttt{c} succeeds then the protocol successfully commits the state updates of all the associated (Root and Segment) transactions,
\item If \texttt{c} fails then the protocol discards the state updates of all the associated transactions.
\end{itemize}

% \textbf{Case Analysis}: Analysed the different ways the protocol can finish.

\textbf{Happy case}: When all transactions go through as planned before the time-out, the Root Transaction commits the updated state on the Root Blockchain, and records the commit message in the Root Event. Because of A1 and A3, the Signalling Transactions are submitted on the Segment blockchains and the provisional states are committed to the respective blockchains. 

\textbf{Failure of a Segment Transaction or Root Transaction}: If any of the Segment Transactions fail, then the corresponding Segment Event records an error. Subsequently, the processing of the Root Transaction reads that error, records an Abort flag in the Root Event. An Abort flag is also recorded in the Root Event if the Segment Transactions are successful but the Root Transaction fails. When Signalling Transactions are submitted with this Root Event showing the Abort flag, the provisional updates on Segment Blockchains are discarded.

%Note that the crashing of an application or it becoming Byzantine is included in this case.

\textbf{Failure of the Start Transaction}: When the processing of a Start Transaction fails, the Start Event is not emitted. Without the Start Event, the Segment Transactions nor the Root Transaction can be processed. The crosschain transaction never gets processed. This is equivalent to a failed crosschain transaction resulting in discarding of provisional states.

\textbf{Timeout}: In this scenario the Start Transaction is successfully processed. However, the Root Transaction is not processed prior to the Crosschain Timeout. Because of A2 the Abort flag is recorded in the Root Event due to timeout when  processing the Root Transaction. Because of A3, we have that the provisional updates on other blockchains are eventually discarded.

\subsection{Liveness} 
\textbf{Claim}: The protocol terminates after a finite number of steps.

\textbf{Proof by contradiction}: We have the Crosschain Timeout value recorded in the Start Event. Suppose the protcol does not terminate. Then the timeout happens. The Start Event is a parameter to the Root Transaction. Because of A2, when the Root Transaction is processed, it emits a Root Event with the Abort flag set. Because of A3, the provisional updates are discarded, albeit the protocol terminates.

Note that there are no centralised entities in this protocol, apart from the validators that sign the events. From A1, and from A4 the nodes becoming Byzantine does not affect the termination of the protocol.

\subsection{Deadlocks and Livelocks}
The liveness property implies that the protocol is deadlock free. However, the protocol is not livelock free. Consider the case of two applications \texttt{A1} and \texttt{A2} submitting two crosschain transactions \texttt{T1} and \texttt{T2}, both wanting to update the same contracts \texttt{C1} and \texttt{C2} on blockchains \texttt{B1} and \texttt{B2} respectively. Application \texttt{A1} has as the target of the Root Transaction contract \texttt{C1} and application \texttt{A2} has as the target of the Root Transaction contract \texttt{C2}. The Segment Transactions will go through, locking contract \texttt{C2} for application \texttt{A1} and contract \texttt{C1} for application \texttt{A2}. However, the Root Transactions to update contracts \texttt{C1} and \texttt{C2} will fail as the contracts would already be locked. Both crosschain transactions will fail. The applications could repeatedly resubmit these transactions, and repeating the same failure situation. Hence, it is possible for the protocol to be in a livelock continuously. However, good software design such as the Router - Item pattern~\cite{robinson2020-cross-perf} will greatly reduce the probability of this.

\subsection{Replay Protection}
A parameter to the Start function is the Crosschain Transaction Identifier. The identifier is emitted in the Start Event. This identifier and the Root Blockchain's identifier are recorded in the Root Blockchain and Segment Blockchains Crosschain Control Contracts to ensure Segment Transactions and Root Transactions are not submitted more than once for the one crosschain call.

%%%%%%%%%%%%%%%%%%%%%%%%%%%%%%%%%%%%%%%%%%%%%%%%%%%%%%%%%%%
\section{Evaluation and Discussion}
\label{sec:method}
\label{sec:perf}
This section describes the gas usage and the latency for crosschain transactions executed using this protocol in three scenarios: read an integer from one blockchain to another and store the value on the first blockchain, write an integer from one blockchain to another and store the value on the second blockchain, and the trade finance application described in Figure~\ref{fig:usage}. 
\ifARXIV
Due to the similarities between the Block Header Transfer and the Block Header Transfer via a Coordinating Blockchain techniques, only the Direct Signing and the Block Header Transfer techniques have been evaluated. The main impact on crosschain transactions of the Block Header Transfer via a Coordinating Blockchain technique when compared to the Block Header Transfer technique is increased latency. 
For comparison purposes, the cost of executing the call execution trees as single blockchain function call is presented.
\fi

Gas usage has been evaluated as it relates to how many transactions per second can be executed. For example, Ethereum MainNet currently has a gas limit per block of 12.5 million~\cite{ethstats}. Ethereum clients used in consortium blockchains can have block gas limits much higher. Latency has been evaluated because it describes how long the protocol takes to complete.

The results were created by running five blockchains, each with a single blockchain node, on a Mac Book Pro 2017 laptop with a 2.9 GHz Quad-Core Intel Core i7 CPU and with 16 GB or RAM. The source code, scripts, and detailed instructions that are required to reproduce the results are available on github.com\cite{gpact_github}.

Latency is presented in terms of numbers of \textit{Finalised Block Periods}. Different blockchains and different consortiums using blockchains have different block confirmation requirements. That is, for blockchains using consensus protocols that offer probabilistic finality, such as PoW, some consortium may require one in a million certainty that a block will not be reverted given certain attacker scenarios, whereas other organisations may require greater certainty. Additionally, some blockchains offer instant finality, where a block is deemed final as soon as it has been published. Some blockchains produce new blocks each second, whereas some blockchains only produce new blocks each ten minutes. To normalise between these chains, a \textit{Finalised Block Period} is defined as the time required for a block to be deemed final. For a consortium blockchain with instant finality and a one second block period, a single \textit{Finalised Block Period} would be one second. For Ethereum MainNet that requires eight block confirmations to offer one in a million probability that a block will remain part of the canonical chain and the blockchain will not reorganise such that the block becomes stale assuming 10\% attacker mining power~\cite{robinson2020-mainnet}, with an average block period of 13 seconds, a single \textit{Finalised Block Period} would be 104 seconds.

\subsection{Gas Usage}

\begin{table}
  \centering
    \begin{tabular}{| l | r | r |}
    \hline
   \multirow{2}{*}{Function} &  \multicolumn{2}{|c|}{Gas Cost}  \\
       \cline{2-3}
                         & \multicolumn{1}{|c|}{Direct Signing} & \multicolumn{1}{|c|}{Block Header} \\ 
                         &                             & \multicolumn{1}{|c|}{Transfer} \\ 
       \hline
Start                              &    51,926    &   51,926 + 62,787  \\
Segment (read)            &   230,032   & 226,976 + 62,787  \\
Root                            &    483,556   & 479,036 + 62,787 \\
       \hline
\ifARXIV
       \multicolumn{3}{|l|}{Single Blockchain Transaction} \\
       \hline
read                              &  \multicolumn{2}{|c|}{29,662}  \\
       \hline
\fi
  \end{tabular}
  \caption{Gas Cost: Crosschain Read}
  \label{table:read}
\end{table}

\begin{table}
  \centering
    \begin{tabular}{| l | r | r |}
    \hline
   \multirow{2}{*}{Function} &  \multicolumn{2}{|c|}{Gas Cost}  \\
       \cline{2-3}
                                        & \multicolumn{1}{|c|}{Direct Signing} & \multicolumn{1}{|c|}{Block Header} \\ 
                                        &                             & \multicolumn{1}{|c|}{Transfer} \\ 
       \hline
Start                                &   52,898  & 52,898  \\
Segment (write)              &  403,616 & 370,685  + 62,787\\
Root                               &  397,295 & 410,737 + 62,787\\
Signalling                       &  182,895 & 218,335 + 62,787\\
       \hline
\ifARXIV
       \multicolumn{3}{|l|}{Single Blockchain Transaction} \\
       \hline
write                             & \multicolumn{2}{|c|}{29,658}  \\
       \hline
\fi
  \end{tabular}
  \caption{Gas Cost: Crosschain Write}
  \label{table:write}
\end{table}

\begin{table}
  \centering
    \begin{tabular}{| l | r | r |}
    \hline
   \multirow{2}{*}{Function} &  \multicolumn{2}{|c|}{Gas Cost}  \\
       \cline{2-3}
                         & \multicolumn{1}{|c|}{Direct Signing} & \multicolumn{1}{|c|}{Block Header} \\ 
                         &                             & \multicolumn{1}{|c|}{Transfer} \\ 
       \hline
Start                               & 72,811 & 72,811  \\
Segment (getPrice)        & 495,049   & 494,397 + 62,787\\
Segment (transfer)         & 735,942  & 735,058 + 62,787\\
Segment (delivery)        & 735,920  & 735,036 + 62,787\\
Segment (shipment)      & 946,590   & 943,631 + 62,787\\
Root (executeTrade)      & 845,493 & 941,560 + 62,787\\
Signalling (transfer)      & 145,439  & 186,691 \\
Signalling (deliver)       & 145,439  & 186,691 \\
       \hline
\ifARXIV
       \multicolumn{3}{|l|}{Single Blockchain Transaction} \\
       \hline
executeTrade              & \multicolumn{2}{|c|}{96,725}  \\
       \hline
\fi
  \end{tabular}
  \caption{Gas Cost: Trade Finance Application}
  \label{table:trade}
\end{table}

Tables~\ref{table:read}, \ref{table:write}, and \ref{table:trade} shows the gas usage for the protocol for various scenarios. For all scenarios, when using the Block Header Transfer event transfer technique, in addition to executing the Start, Segment, or Root function, the signed block header needs to be transferred to blockchains that consume the events. The cost of this call is \texttt{62,787} gas on each of the consuming blockchains. 

The Start function gas usage depends on the base transaction fee \texttt{21,000} gas, the cost of storing the crosschain transaction id \texttt{20,000} gas, and the size of the call execution tree that is emitted as part of the Start Event. The call execution tree for the Read and Write scenarios are similar size and hence the gas cost is similar. The call execution tree for the Trade Finance scenario is larger, and hence the gas cost is more for this scenario.

The Segment function gas usage has many factors. The main factors driving gas usage are the size of the call execution tree, the number of locked contracts, and the number of signers. The call execution tree is stored to storage during the life of the Segment function to have the expected call execution tree available when the code makes outgoing crosschain calls. The larger the call execution tree, the more gas used. The address of each locked contract needs to be stored when a contract is locked so that the list of locked contracts can be emitted in Segment Events. The Read scenario does not lock a contract, whereas the Write scenario does. The Trade Finance scenario has a locked contract on the Finance and Logistics blockchains. 

Similar to the Segment function, there are multiple factors driving the gas usage for the Root function. The main factors driving gas usage are the size of the call execution tree, whether lockable contracts are used on the root blockchain, and the number of signers. The Write scenarios does not use lockable storage on the Root Blockchain whereas the Read and the Trade Finance scenarios do. This factor drives increased gas cost for these scenarios. 

The Read scenario does not have a Signalling call as it does not use any lockable contracts on the blockchain being read from. The Write and the Trade Finance scenario have Signalling calls as they use lockable storage on Segment blockchains. In particular, the Signalling call is used on the Finance and Logistics blockchains to unlock the lockable storage on those blockchains. 

\ifARXIV
For all scenarios and both event transfer techniques, the gas required to execute the call execution tree is significantly greater than required by a scenario when all of the contracts were on the same blockchain and a single blockchain call were used. This indicates that effort should be put into optimising gas usage for the protocol implementation and that crosschain function calls should be used only when necessary.
\fi

It should be noted that the gas usage is based on gas costs for the Constantinople fork. See~\cite{gpact_github} for the precise configuration of Hyperledger Besu to replicate these gas costs.

\subsection{Latency}

\begin{table}
  \centering
    \begin{tabular}{| l | c | c | c | c |}
    \hline
   \multirow{2}{*}{Scenario} &  \multicolumn{2}{|c|}{Serial Execution} &  \multicolumn{2}{|c|}{Parallel Execution} \\
       \cline{2-5}
                         & Direct    & Block      & Direct     & Block     \\ 
                         & Signing & Header    & Signing  & Header  \\ 
                         &              & Transfer  &               & Transfer   \\ 
       \hline
Read                &    3        &   5           & 3             & 5  \\
Write                &    4        &   7          & 4             & 7  \\
Trade Finance &         7       &   13     &   5          &    9\\
       \hline
  \end{tabular}
  \caption{Crosschain Transaction Latency measured in Finalised Block Periods}
  \label{table:latency}
\end{table}

Table~\ref{table:latency} shows the latency of the three scenarios when executed with Direct Signing and Block Header Transfer event transfer techniques when using the Serial and Parallel Execution Engines, in terms of Finalised Block Periods. There is no latency difference when using the Serial or Parallel Execution Engines when the call execution tree is simple, as is the case for the Read and Write scenarios. As expected, when there are multiple calls from the one function, as is the case in the Trade Finance scenario, using a Parallel execution engine is advantageous. Using Block Header Transfer tends to almost double the latency compared to Direct Signing consensus. The reason for this is that extra transactions are needed to add the Block Header to the blockchain. This extra latency counter-balances the advantage of Block Header Transfer consensus, where validators only need to sign one block header per block, rather than each crosschain transaction.

%%%%%%%%%%%%%%%%%%%%%%%%%%%%%%%%%%%%%%%%%%%%%%%
\ifARXIV

\section{Crosschain for Heterogeneous Blockchains}
Two changes are required to allow this protocol to work across non-Ethereum blockchains in addition Ethereum blockchains. The receiving blockchain's Crosschain Control Contract needs to be able to translate from Ethereum encoded function calls to their native function call format. Validators for the non-Ethereum blockchain need to translate event data to the Ethereum format and sign the translated event data.

\fi

%%%%%%%%%%%%%%%%%%%%%%%%%%%%%%%%%%%%%%%%%%%%%%%
\section{Limitations}
The protocol has three known limitations: the call tree must be deterministic, Ethereum Events need to be trusted, and applications need to be designed to not be susceptible to Front Running Attacks\cite{front-running}. For example, if a Root function passes a parameter to a Segment function that depends on the block's timestamp, the simulation of the call tree will not be able to predict the parameter value. If Ethereum Events are not trusted, then return values from Segment Events and commit or discard values in Root Events can not be trusted. 

% Front Running Attacks involve attackers observing transactions being submitted to a network and then attempt to submit their own transaction and have it added to a block ahead of the observed transaction.

%%%%%%%%%%%%%%%%%%%%%%%%%%%%%%%%%%%%%%%%%%%%%%%
\section{Conclusion}
This paper introduces and empirically examines the General Purpose Atomic Crosschain Transactions protocol, a protocol that provides atomic, synchronous, inter-contract function calls across blockchains. This protocol is the first protocol that offers general purpose crosschain function calls without any changes to the existing blockchain platforms. It works on public permissionless Ethereum blockchains and consortium Ethereum blockchains. 

The performance of the protocol depends on the size of the crosschain call execution tree. Therefore developers utilising this technology need to understand their crosschain functions' call execution tree to optimise system performance. They need to determine important high value transactions that need to be crosschain to minimise the impact on performance. 

There is much future work to be undertaken. The authors expect to be able to significantly optimise the gas usage of the source code. Improved execution engines are expected to improve the latency for call execution trees that can not be executed by the simplistic Parallel Execution Engine. Finally, implementing the system on non-Ethereum blockchains would allow crosschain calls across heterogeneous blockchains. 
\ifARXIV
The freely available codebase that has been developed as part of this research can be used for exploring these possibilities.
\fi

%\ifCLASSOPTIONcompsoc
%\else
  \section*{Acknowledgment}
%\fi

This research has been undertaken whilst we have been employed full-time at ConsenSys Software. Peter acknowledges the support of his PhD co-supervisors Dr Marius Portmann and Dr David Hyland-Wood. We acknowledge the researchers who have helped develop the Atomic Crosschain Transactions for Ethereum Private Sidechains technology upon which this protocol is based: Dr Sandra Johnson, Dr David Hyland-Wood, Roberto Saltini, Horacio Mijail Anton Quiles, John Brainard, and Zhenyang Shi. We acknowledge Dr Catherine Jones for her thoughtful review comments and suggestions. 

\bibliographystyle{IEEEtran}
\bibliography{IEEEabrv,ref}

\end{document}